\documentclass[12pt]{article}

\font\grande=cmr9.5 scaled \magstep4
\font\medio=cmr9.5 scaled \magstep2
\outer\def\beginsection#1\par{\medbreak\bigskip
      \message{#1}\leftline{\bf#1}\nobreak\medskip
\vskip-\parskip
      \noindent}
\usepackage{graphicx} 
\begin{document}
\bibliographystyle {unsrt}

\titlepage

\begin{flushright}
CERN-PH-TH/2010-176
\end{flushright}

\vspace{15mm}
\begin{center}
{\grande Secondary graviton spectra and waterfall-like fields}\\
\vspace{1.5cm}
 Massimo Giovannini 
 \footnote{Electronic address: massimo.giovannini@cern.ch} \\
\vspace{1cm}
{{\sl Department of Physics, 
Theory Division, CERN, 1211 Geneva 23, Switzerland }}\\
\vspace{0.5cm}
{{\sl INFN, Section of Milan-Bicocca, 20126 Milan, Italy}}
\vspace*{2cm}

\end{center}

\vskip 1cm
\centerline{\medio  Abstract}
\vskip 1cm
The secondary  spectra of the gravitons induced by a waterfall-like field are computed and the general bounds on the spectral energy density of the tensor modes of the geometry are translated into explicit constraints 
on the amplitude and slope of the waterfall spectrum. 
The obtained results are compared with the primary gravitational wave spectra of the concordance model and of its neighboring extensions as well as with the direct Ligo/Virgo bounds on stochastic backgrounds of relic gravitons. Space-borne interferometers (such as Lisa, Bbo, Decigo) seem to be less relevant but their potential implications are briefly outlined.
\noindent

\vspace{5mm}
\vfill
\newpage
\renewcommand{\theequation}{1.\arabic{equation}}
\setcounter{equation}{0}
\section{Motivations}
\label{sec1}
There are situations, in the early Universe,  where waterfall-like fields arising in hybrid inflation \cite{linde} are amplified with spectral slopes which are even steeper than the ones characterizing the vacuum fluctuations. In the latter case the inhomogeneities of the scalar field induce a secondary relic graviton spectrum as opposed to the primary spectra induced directly though the coupling of the tensor modes of the geometry to the the space-time curvature. Provided the slopes of the secondary spectra are sufficiently steep, they can dominate the energy density of the produced gravitons at high frequencies. In this paper the secondary tensor spectra will be computed within a sufficiently general parametrization so that the obtained results can be of wider applicability.
Recalling that the two point function of a canonically normalized scalar field is customarily defined as 
\begin{equation}
 \langle \sigma(\vec{k},\tau) \sigma(\vec{p},\tau) \rangle = \frac{2 \pi^2 }{k^3} {\mathcal P}_{\sigma}(k,\tau) \delta^{(3)}(\vec{k} + \vec{p}), 
\label{spect1}
\end{equation}
the power spectrum of $\sigma$ can be parametrized as\begin{equation}
{\mathcal P}_{\sigma}(k,\tau) = A_{\sigma}^2 \biggl(\frac{k}{k_{\mathrm{max}}}\biggr)^{n_{\sigma} -1} {\mathcal F}(k\tau),
\label{spect2}
\end{equation}
where $A_{\sigma}$ has the dimension of an inverse length and $n_{\sigma} $ is the spectral slope. In Eq. (\ref{spect2}) the dimensionless function ${\mathcal F}(k \tau)$ accounts for the time dependence  and the power specctrum is normalized at the comoving wavenumber $k_{\mathrm{max}}$ corresponding to a comoving frequency $\nu_{\max}$ 
\begin{equation}
\nu_{\mathrm{max}}  = \frac{k_{\mathrm{max}}}{2\pi} = 0.417\,\beta \,\biggl(\frac{\epsilon}{0.01}\biggr)^{1/4} 
\biggl(\frac{{\mathcal A}_{\mathcal R}}{2.28 \times 10^{-9}}\biggr)^{1/4}
\biggl(\frac{h_{0}^2 \Omega_{\mathrm{R}0}}{4.15 \times 10^{-5}}\biggr)^{1/4} \,\mathrm{GHz},
\label{spect2a}
\end{equation}
where the fiducial values\footnote{In Eq. (\ref{spect2a}) ${\mathcal A}_{{\mathcal R}}$ denotes the amplitude of the curvature perturbations (see also Eq. (\ref{spect3})); $\Omega_{0X}$ denotes the present 
value of the energy (or matter) density parameter in critical units and for the species $X$.} are the ones corresponding to the WMAP 7yr data alone \cite{wmap7} supplemented by a tensor component  with $r_{\mathrm{T}} = 16\, \epsilon < 0.36$ (with $\epsilon= - \dot{H}/H^2$ the standard slow-roll parameter); the parameter $\beta$ appearing in Eq. (\ref{spect2a}) depends upon the width of the transition between the inflationary phase and the subsequent radiation dominated phase. In the notations  of Eqs. (\ref{spect1}) and (\ref{spect2}) the scale-invariant limit corresponds to $n_{\sigma} =1$. The case $n_{\sigma}= 3$ characterizes the slope of quantum (vacuum) fluctuations. 
If $n_{\sigma} > 3$ the spectral slope is steeper than  in the case of vacuum fluctuations. The amplified spectrum characterizing the waterfall field in hybrid inflation leads, according to recent analyses \cite{w1}, to $n_{\sigma} \simeq 4$. The curvature perturbations at the comoving scale $k_{\mathrm{p}} = 0.002\, \mathrm{Mpc}^{-1}$ are characterized by an amplitude ${\mathcal A}_{{\mathcal R}}$
\begin{equation}
\langle {\mathcal R}(\vec{k},\tau) {\mathcal R}(\vec{p},\tau) \rangle = \frac{2 \pi^2}{k^3} 
{\mathcal P}_{{\mathcal R}}(k) \delta^{(3)}(\vec{k} + \vec{p}), \qquad {\mathcal P}_{{\mathcal R}}(k) = {\mathcal A}_{{\mathcal R}} \biggl(\frac{k}{k_{\mathrm{p}}}\biggr)^{n_{\mathrm{s}}-1}.
\label{spect3}
\end{equation}
In the notation of Eq. (\ref{spect3}) the curvature perturbations 
induced by the waterfall field can be estimated for scales comparable with $k_{\mathrm{p}}$ and the results are such that ${\mathcal A}_{{\mathcal R}} \simeq {\mathcal O}( 10^{-54})$ \cite{w1}. The latter figure could be 
compared, for instance, with the amplitude 
of the curvature perturbations induced by the adiabatic mode are   ${\mathcal A}_{{\mathcal R}} = (2.43 \pm 0.11)\times 10^{-10} $ with $n_{\mathrm{s}} = 0.963 \pm 0.014$ according to the WMAP 7yr data alone analyzed in the light of the $\Lambda$CDM paradigm and without tensors \cite{wmap7}.

The smallness of the induced curvature perturbations at the scale 
 $k_{\mathrm{p}}$ (corresponding to a typical multipole $\ell_{\mathrm{p}} \simeq 30$) guarantees the absence of modifications in the temperature and polarization anisotropies for all the multipoles 
relevant in CMB studies. The smallness of the scalar fluctuations of the geometry in the frequency range probed by CMB physics \cite{w1} does not exclude the possibility of a sharp increase of the spectral energy density of the tensor modes of the geometry at higher frequencies.  Indeed, as it is well known, the typical frequencies probed by CMB physics correspond 
to low-frequency gravitons in the aHz range while $\nu_{\mathrm{max}}$ is 
rather in the GHz range\footnote{Recall that $1 \, \mathrm{aHz} = 10^{-18} 
\mathrm{Hz}$ while $1\, \mathrm{GHz} = 10^{9} \, \mathrm{Hz}$.}.
In the concordance model \footnote{The concordance 
model will be taken as a synonym of the $\Lambda$CDM scenario where 
$\Lambda$ stands for the dark energy component and CDM for the 
cold dark matter contribution.} the tensor modes are, strictly speaking, absent. It is however possible to include them by adding, in the simplest 
situation, only one parameter (i.e. $r_{\mathrm{T}}$, already introduced 
after Eq. (\ref{spect2a}))
which controls, simultaneously, the tensor spectral amplitude and the tensor spectral slope. For frequencies in the GHz range and in the minimal tensor extension of the $\Lambda$CDM model the spectral energy density 
of the relic gravitons can be easily ${\mathcal O}(10^{-16})$ 
in units of the critical energy density and for $r_{\mathrm{T}} \simeq 
{\mathcal O}(10^{-3})$.  Can this figure change (even dramatically) when the 
secondary contribution to the relic graviton background is consistently taken 
into account? How large can the combined spectral energy density 
be at high frequencies? How does the resulting graviton spectral energy density compare with the other potential signals arising 
in slightly different extensions of the $\Lambda$CDM paradigm?
These are few of the questions which will be specifically addressed 
hereunder. The purpose of section \ref{sec2} is therefore to compute the slope and amplitude of the secondary graviton spectra. In section \ref{sec3} the results of \ref{sec2} will be used to ascertain wether and in which region of the parameter space the current bounds on the stochastic backgrounds of relic gravitons are adequately satisfied. In section \ref{sec4} the secondary graviton spectra are compared with the 
primary signal induced directly by the amplified quantum fluctuations of the tensor modes of the geometry. Section \ref{sec5} contains the concluding considerations. 

\renewcommand{\theequation}{2.\arabic{equation}}
\setcounter{equation}{0}
\section{Secondary spectra of relic gravitons}
\label{sec2}
In  hybrid inflationary models \cite{linde} the scalar potential can be written in terms of the slowly rolling inflaton field $\varphi$ and in terms of the waterfall field $\sigma$ endowed with a mexican hat potential and directly coupled to the inflaton:
\begin{equation}
W(\varphi,\sigma) = \frac{1}{4} ( 2 m_{\sigma}^2 - \sqrt{\lambda}  \sigma^2)^2 + \frac{m_{\varphi}^2}{2} \varphi^2 + \frac{g^2}{2} \varphi^2 \sigma^2.
\label{WT1}
\end{equation}
The true vacuum of $\sigma$ arises for $\sigma^2 = 2 m_{\sigma}^2/\sqrt{\lambda}$ and $\varphi=0$ while the false vacuum occurs for $\sigma=0$. The spectrum of $\sigma$ close to the background solution $\sigma=0$ falls into the class described by Eqs. (\ref{spect1}) and (\ref{spect2}) \cite{w1}.  At $\varphi_{\mathrm{c}}^2 = 2 \sqrt{\lambda} m_{\sigma}^2/ g^2$ the waterfall field becomes almost suddenly massless, and the 
(p)reheating phase start. To keep the discussion reasonably general, the case of a waterfall-like degree of freedom characterized by a steep power spectrum\footnote{Even if the case $n_{\sigma} > 2$ is the one directly related to the waterfall spectra, the value of
the spectral index will be kept generic to allow for a wider applicability 
of the obtained results.} $n_{\sigma} > 2$ will be considered. 
The latter assumptions are consistent with the sudden approximation 
where the transition between the inflationary and the radiation-dominated phase is almost instantaneous. In the case of the sudden approximation the parameter $\beta$ of Eq. (\ref{spect2a}) is exactly equal to $1$ but it can
become different from $1$ if the transition to radiation is not sudden as 
computed in explicit numerical examples \cite{mgs}. The tachyonic instability leading to the graviton spectra discussed in the present paper might also arise, under certain conditions, in inflationary models motivated by supergravity \cite{sugra}. In this sense waterfall-like field (not necessarily arising in standard hybrid models) could be touched by some of the present considerations which will be kept, in what follows, as general as possible.
To compute the spectral energy density of the relic gravitons the relevant quantity is the anisotropic stress of the field $\sigma$, i.e. 
\begin{equation}
\Pi(\vec{x},\tau) = \partial_{i} \sigma \partial_{j} \sigma - \frac{1}{3} \delta_{ij} (\partial_{k}\sigma)^2,
\label{WT2}
\end{equation}
whose projection on the two tensor polarizations determines the spectral energy density 
of the relic gravitons. Consider, therefore, a conformally flat Friedmann-Robertson-Walker (FRW) 
geometry supplemented by its tensor fluctuations $\delta_{\mathrm{t}} g_{ij}$, i.e.
\begin{equation}
\overline{g}_{\mu\nu} = a^2(\tau) \eta_{\mu\nu}, \qquad \delta_{\mathrm{t}} g_{ij}(\tau,\vec{q}) = - a^2(\tau) h_{ij}(\vec{q},\tau).
\label{WT2a} 
\end{equation}
 The two polarizations of $h_{ij}(\vec{q},\tau)$ obey, respectively, 
\begin{eqnarray}
&& h_{\oplus}'' + 2 {\mathcal H} h_{\oplus}' + q^2 h_{\oplus} = 2 \ell_{\mathrm{P}}^2 \Pi_{\oplus}(\vec{q},\tau),
\label{WT3}\\
&& h_{\otimes}'' + 2 {\mathcal H} h_{\otimes}' + q^2 h_{\otimes} = 2 \ell_{\mathrm{P}}^2 \Pi_{\otimes}(\vec{q},\tau),
\label{WT4}
\end{eqnarray}
where the prime denotes a derivation with respect to the conformal time coordinate $\tau$ and ${\mathcal H} = a'/a$. In  Eqs. (\ref{WT3}) and (\ref{WT4}) the Planck length is defined as 
\begin{equation}
\ell_{\mathrm{P}}= \sqrt{8 \pi G} = 1/\overline{M}_{\mathrm{P}},
\end{equation}
while, always, in Eqs. (\ref{WT3}) and (\ref{WT4})
$\Pi_{\oplus}(\vec{q},\tau)$ and $\Pi_{\otimes}(\vec{q},\tau)$ denote the projections of the anisotropic stress on the standard basis of the tensor polarizations, i.e. 
\begin{eqnarray} 
&& \Pi_{\oplus}(\vec{q},\tau)= \frac{1}{2 (2\pi)^{3/2}} \int d^{3} k \,\,\epsilon^{ij}_{\oplus}(\hat{q})\,\, k_{i}\, k_{j}\, \sigma(\vec{k}, \tau) \, \sigma(\vec{q} - \vec{k},\tau),
\label{WT5}\\
&& \Pi_{\otimes}(\vec{q},\tau)= \frac{1}{2 (2\pi)^{3/2}} \int d^{3} k \,\,\epsilon^{ij}_{\otimes}(\hat{q}) \,\, k_{i}\, k_{j}\, \sigma(\vec{k}, \tau) \, \sigma(\vec{q} - \vec{k},\tau),
\label{WT6}
\end{eqnarray}
where $\epsilon_{ij}^{\oplus}(\hat{q})$ and $\epsilon_{ij}^{\otimes}(\hat{q})$ are defined as \footnote{Note that  
$\hat{a}$, $\hat{b}$ and $\hat{q}$ form a triplet of mutually orthogonal unit vectors.}
\begin{equation}
\epsilon_{ij}^{\oplus}(\hat{q})= (\hat{a}_{i} \hat{a}_{j} - \hat{b}_{i} \hat{b}_{j}),\qquad \epsilon_{ij}^{\otimes}(\hat{q}) = (\hat{a}_{i} \hat{b}_{j} + \hat{b}_{i}\hat{a}_{j}).
\label{WT7}
\end{equation}
As already mentioned, it will be assumed that the relaxation of $\sigma$ takes place right at the end of inflation; later on the background geometry is dominated first by radiation and then by matter.  Over the frequencies affected by the radiation-matter transition the leading contribution to the spectral energy density comes from the amplified vacuum fluctuations which left the Hubble radius during inflation and re-entered during the matter-dominated 
epoch (see section \ref{sec4}). Conversely the leading contribution to the hight-frequency part of the spectrum
can be computed from Eqs. (\ref{WT3}) and (\ref{WT4}). More precisely, 
during the radiation-dominated stage of expansion each polarization obeys the equation 
\begin{equation}
h_{(\lambda)}(\vec{q},\tau) = \frac{ 2 \ell_{\mathrm{P}}^2}{q a(\tau)} \int_{0}^{\tau} a(\xi)\, \Pi_{(\lambda)}(\vec{q},\xi) G(q,\tau,\xi) \, d\xi,
\label{WT9}
\end{equation}
where $\lambda= \oplus, \, \otimes$ stands for each of the two polarizations and $G(q,\tau,\xi) = \sin{[q (\tau- \xi)]}$ in the case of a radiation-dominated Universe. To compute the spectral energy density of the relic gravitons 
we need an explicit expression for the energy density. In a FRW background of Eq. (\ref{WT2a}) the action of the tensor modes can be written as (see, for instance, \cite{mgs})
\begin{equation}
S_{\mathrm{gw}} = \frac{1}{8\ell_{\mathrm{P}}^2} \int d^{3} x \, d\tau \sqrt{-\overline{g}} \, \overline{g}^{\alpha\beta}
\partial_{\alpha} h_{ij} \partial_{\beta} h^{ij}.
\label{WT10}
\end{equation}
The energy-momentum tensor derived from Eq. (\ref{WT10}) by varying with respect to $\overline{g}^{\alpha\beta}$, as argued by Ford and Parker \cite{ford}, is the correct quantity to use for the calculation of the spectral energy density:
\begin{equation}
{\mathcal T}_{\mu}^{\nu} = \frac{1}{4\ell_{\mathrm{P}}^2}\biggl[ \partial_{\mu} h_{ij} \partial^{\nu} h^{ij} - \frac{1}{2} \delta_{\mu}^{\nu} 
\overline{g}^{\alpha\beta} \partial_{\alpha} h_{ij} \partial_{\beta} h^{ij} \biggr].
\label{WT11}
\end{equation}
Alternatively one could use the energy-momentum pseudo-tensor \cite{landau1} appropriately generalized to 
curved backgrounds \cite{landau2,landau3}. The energy-momentum pseudo-tensor defined from the second-order variation of the Einstein tensor and the  expression of Eq. (\ref{WT11}) are formally slightly different and 
their peruse in the context of stochastic backgrounds of relic gravitons 
has been carefully gauged in \cite{mgd1,mgd2} (see also \cite{mgs}): the two computational strategies  lead exactly to the same spectral energy density for wavelengths shorter than the Hubble radius 
at each corresponding epoch (i.e. $q \tau \gg 1$) and to quantitatively compatible results (within a 
factor $2$) in the opposite limit. For wavelengths shorter than the Hubble radius the relation between the spectral energy density (in critical units) and the power spectrum is given by\footnote{In the present paper $\ln{}$ denotes 
the natural logarithm while $\log{}$ denotes the common logarithm.} \cite{mgs}
\begin{equation}
\Omega_{\mathrm{GW}}(q,\tau) = \frac{1}{\rho_{\mathrm{crit}}} \frac{d \rho_{\mathrm{GW}}}{d \ln{q}} = 
\frac{q^2}{12 {\mathcal H}^2} P_{\mathrm{T}}(q,\tau),
\label{WT12}
\end{equation}
where $\rho_{\mathrm{crit}} = 3 {\mathcal H}^2/(a^2\, \ell_{\mathrm{P}}^2)$ is the critical energy density and where 
$P_{\mathrm{T}}(q,\tau)$ is the power spectrum satisfying 
\begin{equation}
\langle h_{ij}(\vec{x},\tau) h^{ij}(\vec{x} + \vec{r}, \tau) \rangle = \int d\ln{q} {\mathcal P}_{\mathrm{T}}(q,\tau) \frac{\sin{q r}}{q r}.
\label{WT8}
\end{equation}
The direction of propagation of the graviton $\hat{q}$ can be fixed along the$\hat{z}$ axis; consequently 
denoting with $\mu = \cos{\vartheta}$  the explicit form of $\Omega_{\mathrm{GW}}(q,\tau)$
can be written, from Eq. (\ref{WT9}),  as 
\begin{equation}
\Omega_{\mathrm{GW}}(q,\tau) = \frac{q^3}{6 {\mathcal H}^2 a^2} \biggl(\frac{A_{\sigma}}{\overline{M}_{\mathrm{P}}}\biggr)^{4} 
\int_{k_{\mathrm{p}}}^{k_{\mathrm{max}}} k\,d k \int_{-1}^{1} d\mu\, (1- \mu^2)^2 \, P_{\sigma}(k) 
\frac{P_{\sigma}(|\vec{q}- \vec{k}|)}{|\vec{q} - \vec{k}|^{5}} {\mathcal I}_{(1)}^2(k, q, \tau)
\label{WT13}
\end{equation}
where 
\begin{equation}
 {\mathcal I}_{(1)}(k, q, \tau) = \int_{0}^{\tau} \frac{d\xi}{\xi} \frac{a(\xi)}{\xi} \sin{[q(\tau-\xi)]} \, \sin{(k \xi)}\,
 \sin{[|\vec{q} - \vec{k}| \xi]}.
 \label{WT14}
 \end{equation}
Equation (\ref{WT14}) follows by inserting, into  Eqs. (\ref{WT5}) and (\ref{WT6}), 
\begin{equation}
\sigma(\vec{k},\tau) \sigma(\vec{q} - k, \tau) = \sigma(\vec{k}) \sigma(\vec{q}-\vec{k}) \frac{\sin{(k \tau)}}{k\tau} \,
\frac{\sin{[|\vec{q} - \vec{k}|\tau]}}{|\vec{q} - \vec{k}| \tau}.
\end{equation}
Denoting with $p$ the combination $|\vec{q} - \vec{k}|$ the integral of Eq. (\ref{WT14}) can be expressed as 
\begin{eqnarray}
 {\mathcal I}_{(1)}(k,q,\tau) &=& \frac{\sin{(q \tau)}}{4} \biggl\{  \mathrm{Ci}[(k - p + q)\tau]- \mathrm{Ci}[(q - p - k)\tau] 
\nonumber\\
&+& \mathrm{Ci}[(q + p - k)\tau] 
- \mathrm{Ci}[(k + p + q)\tau] + \ln{\biggl[ \frac{q^2 - (p + k)^2}{q^2 -(k-p)^2}\biggr]}\biggr\}
\nonumber\\
&+& \frac{\cos{(q \tau)}}{4}\biggl\{\mathrm{Si}[(k + p + q)\tau] - \mathrm{Si}[( p + q -k)\tau] 
\nonumber\\
&-& \mathrm{Si}[(k - p +q)\tau] - \mathrm{Si}[(k + p -q)\tau]\biggr\},
\label{WT15}
\end{eqnarray}
where, with standard notations \cite{abr,tric},  
\begin{equation}
\mathrm{Ci}(z) = - \int_{z}^{\infty} \frac{\cos{t}}{t} dt,\qquad \mathrm{Si}(z) =  \int_{z}^{\infty} \frac{\sin{t}}{t} dt.
\label{WT16}
\end{equation}
The small and large argument limit of $\mathrm{Ci}(z)$ and $\mathrm{Si}(z)$ is given, respectively, by
\begin{eqnarray}
&&\mathrm{Ci}(z) = \gamma + \ln{z}  + {\mathcal O}(z^2),\qquad \mathrm{Si}(z) = z  + {\mathcal O}(z^3),
\nonumber\\
&& \mathrm{Ci}(z) = \frac{\sin{z}}{z} - \frac{\cos{z}}{z^2} + {\mathcal O}(z^{-3}), \qquad \mathrm{Si}(z) = \frac{\pi}{2} - \frac{\cos{z}}{z} + {\mathcal O}(z^{-2}).
\label{WT17}
\end{eqnarray}
Note that $z  \gg 1$ all the wavelengths 
are shorter than the Hubble radius and, in this limit 
\begin{eqnarray}
&&{\mathcal I}_{(1)}(k,q,\tau) = - \frac{\pi}{4} \cos{q \tau} + \frac{\sin{q\tau}}{4 (q\,\tau)} \biggl\{ - \frac{4 k q^2 p}{[(q + k)^2 - p^2][(q-k)^2 -p^2]} 
\nonumber\\
&& + q \ln{\biggl[\frac{q^2 - (k + p)^2}{q^2 - (k - p)^2}\biggr]}\biggr\} + {\mathcal O}\biggl(\frac{1}{q^2 \tau^2}\biggr).
\label{WT18}
\end{eqnarray}
The leading contribution to Eq. (\ref{WT13}) is given by the first term (proportional to $\cos{q\tau}$) while the other terms are suppressed as $q \tau >1$.   The oscillations inside the Hubble radius  are typically averaged in the final result by replacing $\cos^2{q\tau} \to 1/2$ (see, e.g. \cite{efs,stein}). Following the same procedure, the final expression for the spectral energy density of the relic gravitons is given by:
\begin{equation}
\Omega_{\mathrm{GW}}(q,\tau) = \frac{\pi^2}{192} \, \biggl(\frac{A_{\sigma}}{\overline{M}_{\mathrm{P}}}\biggr)^{4}  \biggl(\frac{q}{q_{\mathrm{max}}}\biggr)^3 \int_{k_{\mathrm{p}}}^{k_{\mathrm{max}}} \biggl(\frac{k}{k_{\mathrm{max}}}\biggr)^{n_{\sigma}}  {\mathcal I}_{(2)}(q/q_{\mathrm{max}},k/k_{\mathrm{max}}, n_{\sigma})\,\, d (k/k_{\mathrm{max}}),
\label{WT19}
\end{equation}
where, defining $x= k/k_{\mathrm{max}}$ and $y= q/q_{\mathrm{max}}$,  
\begin{equation}
{\mathcal I}_{(2)}(x,y,n_{\sigma}) = \int_{-1}^{1} (1 - \mu^2)^2 [ y^2 + x^2 - 2 x y \mu]^{(n_{\sigma} -6)/2}\, d\mu.
\label{WT20}
\end{equation}
The integrals of Eq. (\ref{WT19}) and (\ref{WT20}) can be performed numerically and, in some approximation, also analytically; the final result can be written as 
\begin{equation}
\Omega_{\mathrm{GW}}(q,\tau) = \frac{\pi^2}{180} \biggl(\frac{A_{\sigma}}{\overline{M}_{\mathrm{P}}}\biggr)^{4} \biggl(\frac{q}{q_{\mathrm{max}}}\biggr)^{2 (n_{\sigma} -1)} \biggl[ \frac{6 -n_{\sigma}}{(n_{\sigma} + 1) ( 5 - 2 n_{\sigma})} +  \frac{1}{2n_{\sigma} - 5} \biggl(\frac{q}{q_{\mathrm{max}}}\biggr)^{5 - 2 n_{\sigma}}\biggr].
\label{WT22}
\end{equation}
Let us now pause for a moment and discuss the approximations used 
to derive Eq. (\ref{WT22}). The first remark is that the integral of Eq. (\ref{WT20}) 
seems to lead to a logarithmic divergence in the case $n_{\sigma} =4$. Even in the 
case $n_{\sigma} =4$,  Eq. (\ref{WT22}) is in very good quantitative agreement with 
the numerical result. The rationale for this occurrence is that Eq. (\ref{WT19}) can be explicitly written, in the case $n_{\sigma} =4$, as  
\begin{eqnarray}
 \Omega_{\mathrm{GW}}(y,\tau) &=& \frac{\pi^2}{192} \biggl(\frac{A_{\sigma}}{\overline{M}_{\mathrm{P}}}\biggr)^{4} G(y),
\label{WT21a}\\
G(y) &=& \frac{1}{96 y^2} \int_{y_{\mathrm{p}}}^{1}\frac{d x}{x} \biggl\{ 3 ( x^2 - y^2)^4 - 4 x y (x^2 + y^2) (3 x^4 - 14 x^2 y^2 + 3 y^4) 
\nonumber\\
&+& 6 (x^2 - y^2)^4\ln{\biggl[\frac{|x + y|}{|x - y|}\biggr]} \biggr\},
\label{WT21b}
\end{eqnarray}
and where $y_{\mathrm{p}} = k_{\mathrm{p}}/k_{\mathrm{max}} = \nu_{\mathrm{p}}/\nu_{\mathrm{max}}$.The integral appearing in ${\mathcal G}(y)$ can be easily performed numerically by recalling that 
\begin{equation}
y_{\mathrm{p}} = 7.41 \times 10^{-27} \frac{1}{\beta} \biggl(\frac{\epsilon}{0.01}\biggr)^{-1/4} 
\biggl(\frac{{\mathcal A}_{\mathcal R}}{2.28 \times 10^{-9}}\biggr)^{-1/4}
\biggl(\frac{h_{0}^2 \Omega_{\mathrm{R}0}}{4.15 \times 10^{-5}}\biggr)^{- 1/4}.
\label{WT21c}
\end{equation}
To obtain a quantitatively accurate expression for the integral it is useful to write 
$G(y)$ as 
\begin{equation}
G(y) = \frac{y^{6}}{96}\biggl[ \int_{y_{\mathrm{p}}}^{y} \frac{d\,x}{x} \, F(u) +
\int_{y}^{1} \frac{d\,x}{x} \frac{F(s)}{s^{8}} \biggr], 
\label{WT21d}
\end{equation}
where $u = x/y$ and $s= y/x$ while for a generic variable $t$ the function $F(t)$ 
is defined as 
\begin{equation}
F(t) = 4 t (1 + t^2)  [ 14 t^2 - 3 ( 1 +t^4)] - 6 (t^2 -1)^4 
\ln{\biggl(\frac{|1\,-\,t|}{|1\,+\,t|}\biggr)}.
\label{WT21e}
\end{equation}
Since $u<1$ for $ y_{\mathrm{p}} < x < y$ and $s< 1$ for $y < x < 1$, $G(y)$ 
can be evaluated by expanding $F(t)$ in the two domains and by keeping 
the first few terms of the expansion and by integrating the obtained result. 
In the limit $y_{\mathrm{p}} \to 0$ the final result for 
$\Omega_{\mathrm{GW}}(q,\tau)$ coincides with the expression given in Eq. 
(\ref{WT22}) for $n_{\sigma} =4$. 

The approximation scheme leading to the result of Eq. (\ref{WT22}) 
is rather accurate both for the present purposes and in more general terms. 
To scrutinize more carefully the latter statement 
the approximate results based on Eq. (\ref{WT22}) will be compared with the explicit numerical integration in various situations. For this purpose it is appropriate to generalize $G(y)$ to the case where $n_{\sigma} \neq 4$:
\begin{equation}
G(y,n_{\sigma}) = y^3 \int_{y_{\mathrm{p}}}^{1} x^{n_{\sigma}} {\mathcal I}_{(2)}(x, y, n_{\sigma}),
\label{WTT1}
\end{equation}
which reduces to Eq. (\ref{WT21b}) when $n_{\sigma}=4$ as it can be easily verified with some 
algebra. The integral of Eq. (\ref{WTT1}) is simply regularized in the ultraviolet by selecting a cut-off 
and this is the rationale for the upper limit of integration. 
\begin{figure}[!ht]
\centering
\includegraphics[height=6cm]{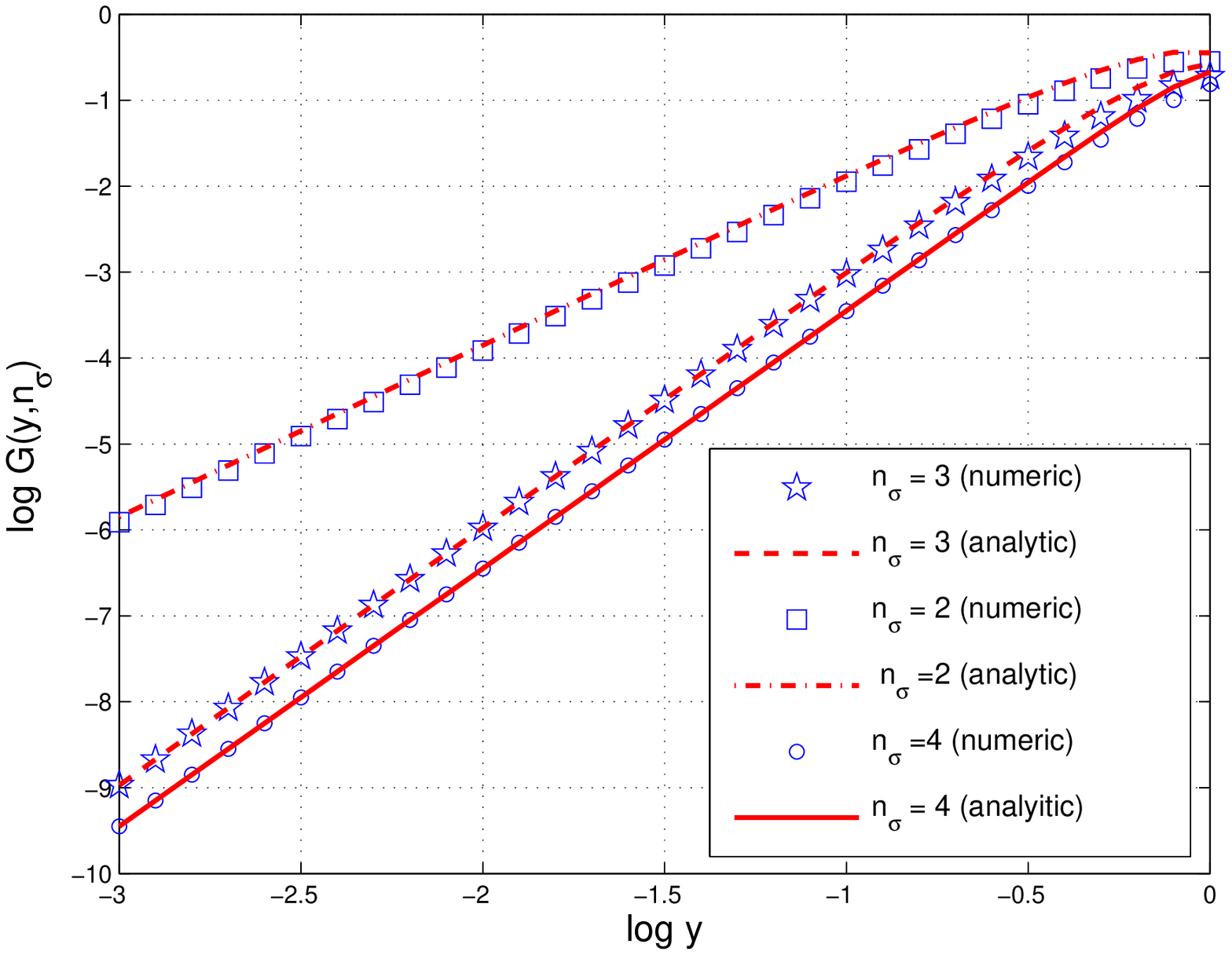}
\includegraphics[height=6cm]{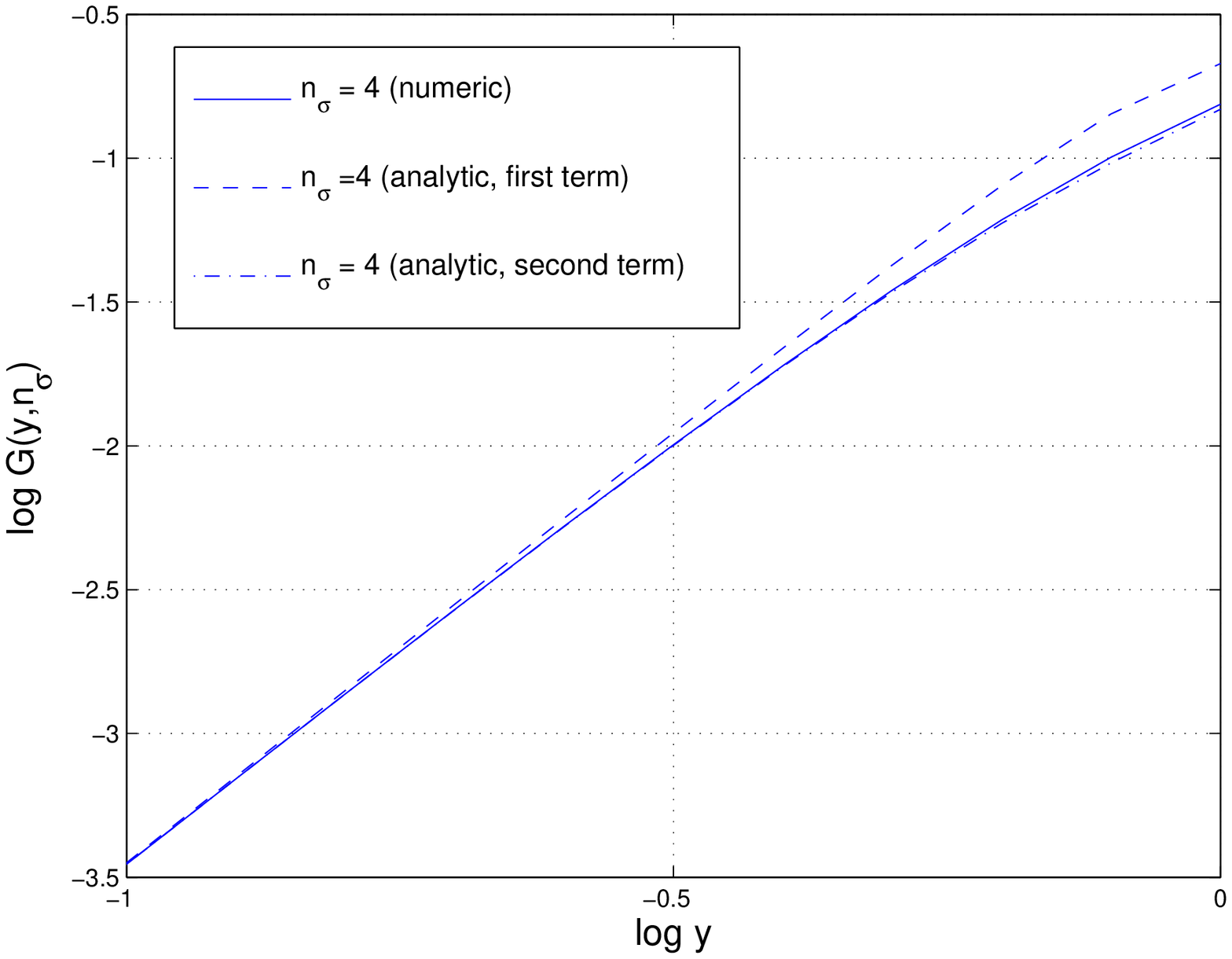}
\caption[a]{The comparison between the numeric and the analytic results for the calculation of $G(y,n_{\sigma})$.}
\label{figure1}      
\end{figure}
The results of the comparison between the numeric evaluation of $G(y,n_{\sigma})$ and the analytic expression
of Eq. (\ref{WT22}) is illustrated in Fig. \ref{figure1}. In the plot at the left the stars, the squares and the naughts 
reproduce, respectively, the numeric results for the different values of $n_{\sigma}$. The lines reproduce instead 
the analytic result obtained on the basis of Eq. (\ref{WT22}). In the plot at the right the results for the case 
$n_{\sigma}=4$ are shown in greater detail for $y$ close to $y=1$. The full line of the right plot shows the 
numeric result while the dashed and dot-dashed lines illustrate instead the analytic estimate. More specifically 
the dashed line shows exactly the result of Eq. (\ref{WT22}) while the dot-dashed line 
is based on the improved approximation where the next-to-leading term is kept when expanding $F(u)$ and $F(s)$ in Eq. (\ref{WT21d}), i.e. 
\begin{eqnarray}
G(y,n_{\sigma}) &=& \frac{16}{15} y^{2 (n_{\sigma}-1)}\biggl\{ \biggl[ \frac{n_{\sigma} - 6}{( 2 n_{\sigma} - 5)(n_{\sigma}+1)} + \frac{y^{5 - 2 n_{\sigma}}}{2 n_{\sigma} -5}\biggr]
\nonumber\\
&+& \frac{(n_{\sigma}-1) (n_{\sigma}-6)}{14} \biggl[\frac{n_{\sigma}-10}{(n_{\sigma} +3) ( 2 n_{\sigma} - 7)} + \frac{y^{7 - 2n_{\sigma}}}{2 n_{\sigma}-7}\biggr]\biggr\}.
\label{WTT3}
\end{eqnarray}
The first line in Eq. (\ref{WTT3}) leads directly to the result of Eq. (\ref{WT22}) while the second line represent the correction. The goodness 
of the agreement between the analytic and the numeric results is therefore quantitatively evident from Fig. \ref{figure1}. 
Every approximation can be improved around the transition frequency $k_{\mathrm{max}}$ by considering the  possible exponential suppression of the modes 
$k> k_{\mathrm{max}}$ as implied, in the case of gravitational particle production, by the so-called adiabatic theorem \cite{bir}. Similar agreement is reached when the cut-off is replaced by the exponential tail. It is 
however important to notice (as noticed in a related context \cite{mgs}) that the width of the transition 
between inflation and radiation affects the definition of $\nu_{\mathrm{max}}$ itself so that $1/\beta$ increases 
from $1$ to $6.33$ in the example of \cite{mgs}. These kinds of inaccuracies imply that the exponential 
tail is itself necessarily inaccurate and must be determined in the context of a specific model. Put in other words it is always possible to produce 
more accurate estimates by purportedly invoking a more accurate 
description of the transition regime. Since in this (as well as in similar cases) 
the understanding of the physics is approximate, we shall be content 
of the present approximation scheme which has the advantage 
of being not only accurate but also transparent. 
\renewcommand{\theequation}{3.\arabic{equation}}
\setcounter{equation}{0}
\section{Observational constraints}
\label{sec3}
The results of  Eq. (\ref{WT22}) estimate the secondary contribution to the 
spectral energy density of the relic gravitons whose primary signal 
emerge because of the direct coupling of the tensor modes 
of the geometry to the space-time curvature. The latter coupling 
amplifies an initial vacuum fluctuation during the inflationary quasi-de Sitter 
stage of expansion and the resulting spectral energy 
density can be written as 
\begin{eqnarray}
h_{0}^2 \Omega_{\mathrm{GW}}(\nu,\tau_{0}) &=& {\mathcal N}_{\mathrm{T}}\,  T^2_{\Omega}(\nu/\nu_{\mathrm{eq}}) \,r_{\mathrm{T}} \, \biggl(\frac{\nu}{\nu_{\mathrm{p}}}\biggr)^{n_{\mathrm{T}}} e^{- 2  \frac{\nu}{\nu_{\mathrm{max}}}},
\label{WT23}\\
{\mathcal N}_{\mathrm{T}} &=& 3.94 \times 10^{-15} \biggl(\frac{h_{0}^2 \Omega_{\mathrm{R}0}}{4.15\times 10^{-5}}\biggr) \biggl(\frac{{\mathcal A}_{{\mathcal R}}}{2.28 \times 10^{-9}}\biggr),
\label{WT24}
\end{eqnarray}
where $r_{\mathrm{T}}$ and $n_{\mathrm{T}}$ are defined as 
\begin{equation}
r_{\mathrm{T}} = \frac{{\mathcal P}_{{\mathcal R}}(k_{\mathrm{p}})}{{\mathcal P}_{\mathrm{T}}(k_{\mathrm{p}})}
,\qquad n_{\mathrm{T}} = - 2 \epsilon + \frac{\alpha_{\mathrm{T}}}{2} \ln{(\nu/\nu_{\mathrm{p}})}, \qquad \alpha_{\mathrm{T}} = \frac{r_{\mathrm{T}}}{8}\biggl[ (n_{\mathrm{s}} -1) + \frac{r_{\mathrm{T}}}{8}\biggr]. 
\label{WT24a}
\end{equation}
If $\alpha_{\mathrm{T}} =0$ the tensor spectral index $n_{\mathrm{T}}$ does not depend upon the frequency and this is the case which is, somehow, endorsed when introducing gravitational waves in the minimal tensor extension of the $\Lambda$CDM paradigm \cite{wmap7}. 
If $\alpha_{\mathrm{T}} \neq 0$ the spectral index is said to be running at a rate that depends upon 
the value of the scalar spectral index $n_{\mathrm{s}}$.  Note that, by definition, $r_{\mathrm{T}}$ represents the ratio between the tensor and the 
scalar power spectrum at the scale $k_{\mathrm{p}}$.
In Eq. (\ref{WT23}) $T_{\Omega}(\nu/\nu_{\mathrm{eq}})$ 
denotes the transfer function of the spectral energy density:
\begin{equation}
T_{\Omega}(\nu/\nu_{\mathrm{eq}}) = \sqrt{1 + c_{1}\biggl(\frac{\nu_{\mathrm{eq}}}{\nu}\biggr) + b_{1}\biggl(\frac{\nu_{\mathrm{eq}}}{\nu}\biggr)^2},\qquad c_{1}= 0.5238,\qquad
b_{1}=0.3537,
\label{WT25}
\end{equation}
where 
\begin{equation}
\nu_{\mathrm{eq}} = \frac{k_{\mathrm{eq}}}{2 \pi} = 1.288 \times 10^{-17} \biggl(\frac{h_{0}^2 \Omega_{\mathrm{M}0}}{0.1334}\biggr) \biggl(\frac{h_{0}^2 \Omega_{\mathrm{R}0}}{4.15 \times 10^{-5}}\biggr)^{-1/2}\,\, \mathrm{Hz}.
\label{WT26}
\end{equation}
Equation (\ref{WT25}) has been derived in \cite{mgd1} (see also \cite{mgs,mgd2}) but other authors prefer to work with the transfer 
function of the power spectrum (see, e. g. \cite{efs,stein}). The two procedures are equivalent but the transfer function of the spectral energy density is easier to assess directly in numerical terms by integrating the 
evolution of the mode functions across the matter-ratiation transition 
for a given initial spectrum. Conversely if one oughts to evaluate the 
transfer function for the power spectrum $P_{\mathrm{T}}(\nu,\tau)$, 
the obtained result must then be connected to $\Omega_{\mathrm{GW}}(\nu,\tau)$ and this should be done carefully since the power spectrum
oscillates and the oscillations  must then be  appropriately averaged 
\cite{mgs}. The specific values of the cosmological parameters determined using the WMAP 7yr data alone 
in the light of the vanilla $\Lambda$CDM scenario are: 
\begin{equation}
( \Omega_{\mathrm{b}}, \, \Omega_{\mathrm{c}}, \Omega_{\mathrm{de}},\, h_{0},\,n_{\mathrm{s}},\, \epsilon_{\mathrm{re}}) \equiv 
(0.0449,\, 0.222,\, 0.734,\,0.710,\, 0.963,\,0.088).
\label{Par1}
\end{equation}
If a tensor component is allowed in the analysis 
of the WMAP 7yr data alone the relevant cosmological parameters are determined to be 
\begin{equation}
( \Omega_{\mathrm{b}}, \, \Omega_{\mathrm{c}}, \Omega_{\mathrm{de}},\, h_{0},\,n_{\mathrm{s}},\, \epsilon_{\mathrm{re}}) \equiv 
(0.0430,\, 0.200,\, 0.757,\,0.735,\, 0.982,\,0.091).
\label{Par2}
\end{equation}
In the case of Eq. (\ref{Par1}) the amplitude of the scalar modes is ${\mathcal A}_{{\mathcal R}} = 
(2.43 \pm 0.11) \times 10^{-9}$ while in the case of Eq. (\ref{Par2}) the corresponding values of ${\mathcal A}_{{\mathcal R}}$ and of $r_{\mathrm{T}}$ are given by 
\begin{equation}
{\mathcal A}_{{\mathcal R}} = (2.28 \pm 0.15)\times 10^{-9},\qquad r_{\mathrm{T}} < 0.36 
\label{Par3}
\end{equation}
to $95$ \% confidence level. To avoid confusions it is appropriate 
to spend a word of care on the figures implied by Eqs. (\ref{Par2}) and 
(\ref{Par3}) which have been used in the numeric analysis just 
for sake of accuracy. At the same time the qualitative features 
of the effects discussed here do not change if, for instance, one 
would endorse the parameters drawn from the minimal tensor extension 
of the $\Lambda$CDM paradigm and compared not to the WMAP 7yr 
data release but rather with the WMAP 3yr data release, implying, for instance, ${\mathcal A}_{{\mathcal R}} = 2.1^{+2.2}_{-2.3}\times 10^{-9}$, 
$n_{\mathrm{s}} =0.984$ and  $r_{\mathrm{T}} < 0.65$ (95 \% confidence level).
With the previous caveats the spectral energy density can be written, in explicit terms, as 
\begin{eqnarray}
h_{0}^2 \Omega_{\mathrm{GW}}(\nu,\tau_{0}) &=& T^2_{\Omega}(\nu/\nu_{\mathrm{eq}})\biggl\{ {\mathcal N}_{\mathrm{T}} r_{\mathrm{T}} \biggl(\frac{\nu}{\nu_{\mathrm{p}}}\biggr)^{n_{\mathrm{T}}} e^{- 2  \frac{\nu}{\nu_{\mathrm{max}}}}
\nonumber\\
&+& {\mathcal N}_{\sigma} \biggl(\frac{\nu}{\nu_{\mathrm{max}}}\biggr)^{2 (n_{\sigma} -1)}\biggl[ \frac{6 -n_{\sigma}}{(n_{\sigma} + 1) ( 5 - 2 n_{\sigma})} +  \frac{1}{2n_{\sigma} - 5} \biggl(\frac{\nu}{\nu_{\mathrm{max}}}\biggr)^{5 - 2 n_{\sigma}}\biggr]\biggr\}
\label{WT27}
\end{eqnarray}
where 
\begin{equation}
{\mathcal N}_{\sigma} = 2.27\times 10^{-6} \,\biggl(\frac{h_{0}^2 \Omega_{\mathrm{R}0}}{4.15 \times 10^{-5}}\biggr) \,\biggl(\frac{A_{\sigma}}{\overline{M}_{\mathrm{P}}}\biggr)^4.
\label{WT28}
\end{equation} 
It is interesting to remark that, for a given (fixed) ratio $\nu/\nu_{\mathrm{max}}$
\begin{equation}
\lim_{n_{\sigma} \to 5/2}\biggl[ \frac{n_{\sigma} - 6}{(n_{\sigma} + 1) ( 2 n_{\sigma} - 5 )} +  \frac{1}{2n_{\sigma} - 5} \biggl(\frac{\nu}{\nu_{\mathrm{max}}}\biggr)^{5 - 2 n_{\sigma}}\biggr] = \frac{2}{7} + \ln{(\nu_{\mathrm{max}}/\nu)},
\label{WT28a}
\end{equation}
so that the apparent divergence occurring 
for $n_{\sigma} = 5/2$ in Eq. (\ref{WT27}) just lead to a logarithmic enhancement. The same kind of limit arises 
in the next-to-leading contribution appearing in the second line of Eq. (\ref{WTT3}), i.e. 
\begin{equation}
\lim_{n_{\sigma} \to 7/2}\biggl[\frac{n_{\sigma}-10}{(n_{\sigma} +3) ( 2 n_{\sigma} - 7)} + \frac{1}{2 n_{\sigma}-7}\biggl(\frac{\nu}{\nu_{\mathrm{max}}}\biggr)^{7 - 2 n_{\sigma}}\biggr] = \frac{2}{13} + \ln{(\nu_{\mathrm{max}}/\nu)}.
\label{WT28b}
\end{equation}
Denoting with $\zeta$ the order of the perturbative correction to the leading result of Eq. (\ref{WT28a}), the 
logarithmic enhancement arise for $n_{\sigma} = (5 + 2 \zeta)/2$ and their magnitude is 
\begin{eqnarray} 
&& \lim_{n_{\sigma} \to (5+ 2 \zeta)/2}\biggl[ \frac{n_{\sigma} - 6 - 4 \zeta }{(n_{\sigma} + 1+ 2 \zeta) (2 n_{\sigma} - 5 - 2 \zeta)} 
\nonumber\\
&& +  \frac{1}{2n_{\sigma} - 5 - 2 \zeta} \biggl(\frac{\nu}{\nu_{\mathrm{max}}}\biggr)^{5 + 2\zeta - 2 n_{\sigma}}\biggr] = \frac{2}{7 + 6 \zeta} + \ln{(\nu_{\mathrm{max}}/\nu)}.
\label{WT28c}
\end{eqnarray}

The relic graviton spectra must satisfy various constraints which will also imply, in particular, quantitative bounds on the parameters of the secondary spectra. The spectral energy density of the relic gravitons must be compatible not only with the CMB constraints (bounding, from above, the value of $r_{\mathrm{T}}$) but also with 
 the pulsar timing bound \cite{pulsar1,pulsar2} and the big-bang nucleosynthesis constraints \cite{bbn1,bbn2,bbn3}. The result of Eq. (\ref{WT23}) automatically incorporates the constraint stemming 
 from the CMB observations  encoded in the value of $r_{\mathrm{T}}$. 
  As we move towards larger frequencies the pulsar timing bound demands
\begin{equation}
\Omega(\nu_{\mathrm{pulsar}},\tau_{0}) < 1.9\times10^{-8},\qquad 
\nu_{\mathrm{pulsar}} \simeq \,10\,\mathrm{nHz},
\label{PUL}
\end{equation}
where $\nu_{\mathrm{pulsar}}$ roughly corresponds to the inverse 
of the observation time during which the pulsars timing has been monitored. 

Following the analyses on the stochastic relic graviton backgrounds
\cite{mgb} (see also \cite{assa}) let us define  $\nu_{\mathrm{LV}}= 100$Hz as the Ligo/Virgo
frequency.  In
Ref. \cite{LIGOS1} (see also \cite{LIGOS2,LIGOS3}) the current limits on the presence of an isotropic 
background of relic gravitons have been illustrated. According to the Ligo collaboration 
(see Eq. (19) of Ref. \cite{LIGOS2}) the spectral energy density of a putative 
(isotropic) background of relic gravitons can be parametrized as:
\begin{equation}
\Omega_{\mathrm{GW}}(\nu,\tau_{0}) = \Omega_{\mathrm{GW},\alpha} \biggl(\frac{\nu}{100\,\mathrm{Hz}}\biggr)^{\alpha + 3}.
\label{LIGOpar}
\end{equation}
For the scale-invariant case (i.e. $\alpha= -3$ in eq. (\ref{LIGOpar}))
the Ligo collaboration sets a $90 \%$ upper limit of $1.20\times 10^{-4}$ on 
the amplitude appearing in Eq. (\ref{LIGOpar}), i.e. $\Omega_{\mathrm{GW},-3}$.
Using different sets of data (see \cite{LIGOS1,LIGOS3}) the Ligo collaboration 
manages to improve the bound even by a factor $2$ getting down to 
$6.5\times 10^{-5}$. Recently this result has been improved to 
$6.9\times 10^{-6}$ \cite{LIGO4} and this is the figure used in the 
forthcoming plots when generically referring to the Ligo/Virgo bound.

Assuming that there are some additional relativistic degrees of
freedom, which also have decoupled by the time of electron-positron
annihilation,  or just some additional component $\rho_X$ to the energy
density with a radiation-like equation of state (i.e. $p_{X} = \rho_{X}/3$), the effect on the expansion rate will be the same as that of having
some (perhaps a fractional number of)
additional neutrino species.   Before
electron-positron annihilation we have $\rho_X = (7/8)\Delta N_{\nu} \rho_\gamma$
and after electron-positron annihilation we have
$\rho_X = (7/8) (4/11)^{4/3} \,\Delta N_{\nu} \,\rho_{\gamma} \simeq 0.227\,\Delta N_{\nu} \,\rho_\gamma$. The critical fraction of CMB photons can be directly computed from the value of the CMB temperature and it is given by
$h_{0}^2 \Omega_\gamma \equiv \rho_\gamma/\rho_{\mathrm{crit}} = 2.47\times10^{-5}$.
If the extra energy density component has stayed radiation-like until today,
its ratio to the critical density, $\Omega_X$, is given by
\begin{equation}
h_{0}^2   \Omega_X \equiv h_{0}^2\frac{\rho_X}{\rho_{\mathrm{c}}} = 5.61\times10^{-6}\Delta N_{\nu} 
\biggl(\frac{h_{0}^2 \Omega_{\gamma0}}{2.47 \times 10^{-5}}\biggr).
\end{equation}
If the additional species are relic gravitons, then  \cite{bbn1,bbn2,bbn3}: 
\begin{equation}
h_{0}^2  \int_{\nu_{\mathrm{bbn}}}^{\nu_{\mathrm{max}}}
  \Omega_{{\rm GW}}(\nu,\tau_{0}) d\ln{\nu} = 5.61 \times 10^{-6} \Delta N_{\nu} 
  \biggl(\frac{h_{0}^2 \Omega_{\gamma0}}{2.47 \times 10^{-5}}\biggr),
\label{BBN1}
\end{equation}
where $\nu_{\mathrm{max}}$ is given by Eq. (\ref{spect2a}) and  $\nu_{\mathrm{bbn}}$ 
is given by:
\begin{equation}
\nu_{\mathrm{bbn}} = 
2.252\times 10^{-11} \biggl(\frac{g_{\rho}}{10.75}\biggr)^{1/4} \biggl(\frac{T_{\mathrm{bbn}}}{\,\,\mathrm{MeV}}\biggr) 
\biggl(\frac{h_{0}^2 \Omega_{\mathrm{R}0}}{4.15 \times 10^{-5}}\biggr)^{1/4}\,\,\mathrm{Hz}\simeq 0.01 \,\,\mathrm{nHz},
\label{WT29}
\end{equation}
In Eq. (\ref{WT29}) $g_{\rho}$ denotes the effective number of relativistic degrees of freedom entering the total energy density of the plasma.  While $\nu_{\mathrm{eq}}$  is still close to the aHz, $\nu_{\mathrm{bbn}}$ is rather in the nHz range. 
\begin{figure}[!ht]
\centering
\includegraphics[height=6cm]{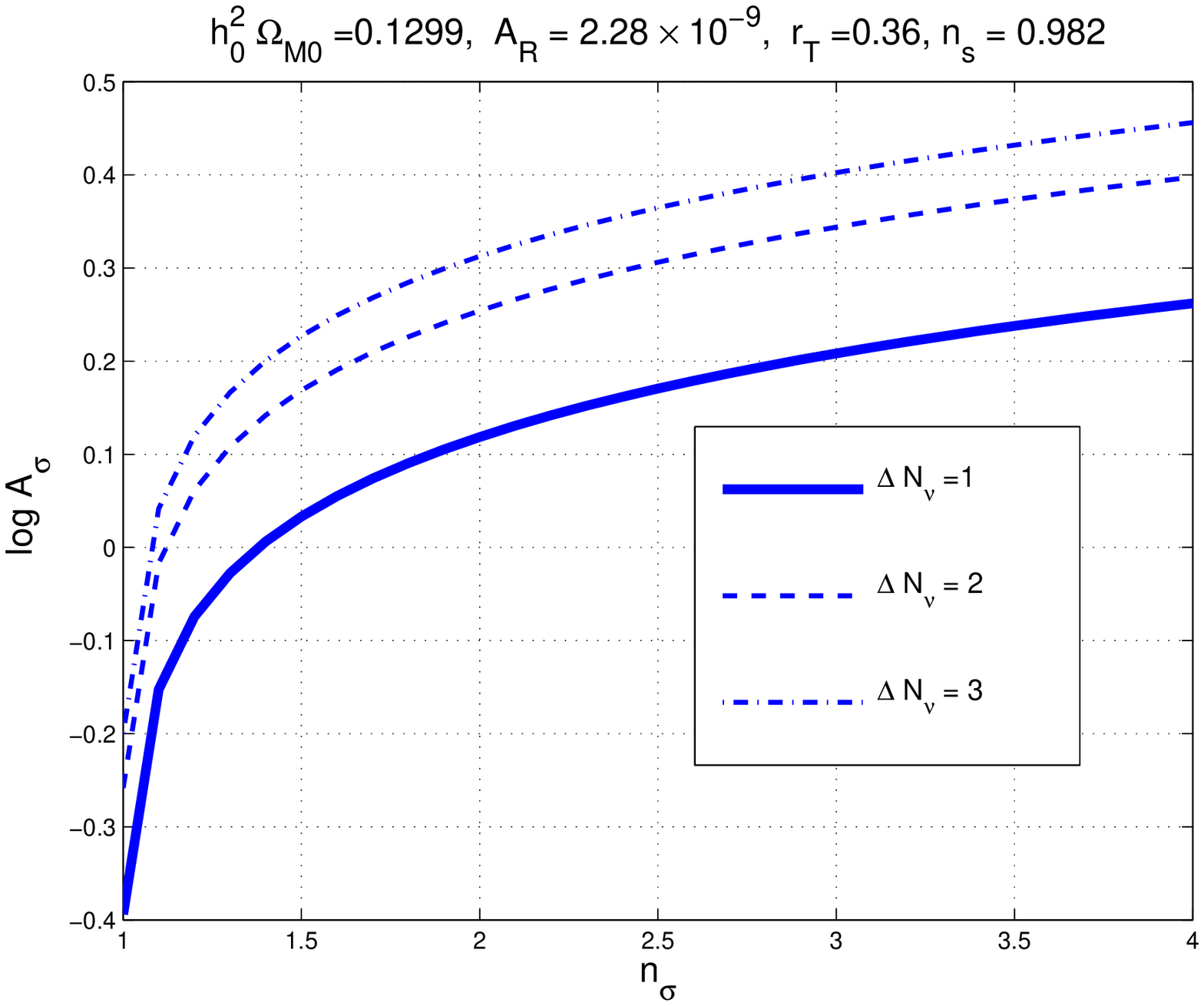}
\includegraphics[height=6cm]{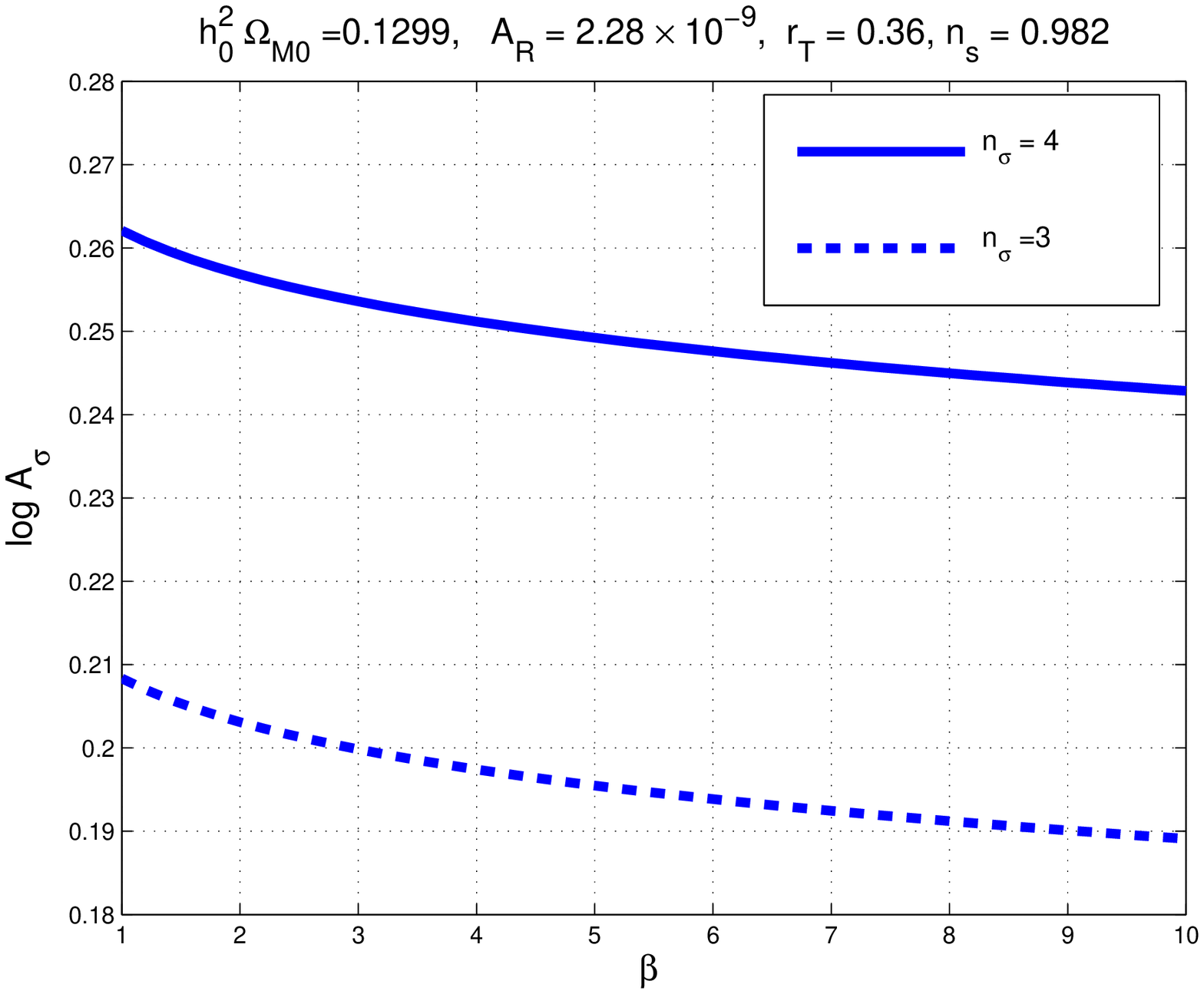}
\caption[a]{The nucleosynthesis limits on the amplitude $A_{\sigma}$ (expressed in units 
of the Planck mass $\overline{M}_{\mathrm{P}}$) for different values of $n_{\sigma}$ and of $\beta$. }
\label{figure2}      
\end{figure}
In Fig. \ref{figure2} the limits on the amplitude $A_{\sigma}$ have 
been illustrated as they arise from Eq. (\ref{BBN1}) and for 
different values of $\Delta N_{\nu}$. In the plot at the left of Fig. \ref{figure2}
the two parameters are $A_{\sigma}$ and $n_{\sigma}$ while in the plot 
at the right, on the horizontal axis, $\beta$ parametrizes the indetermination 
on the maximal frequency or, in a complementary perspective, 
the limitations of the sudden approximation. 
\begin{figure}[!ht]
\centering
\includegraphics[height=6cm]{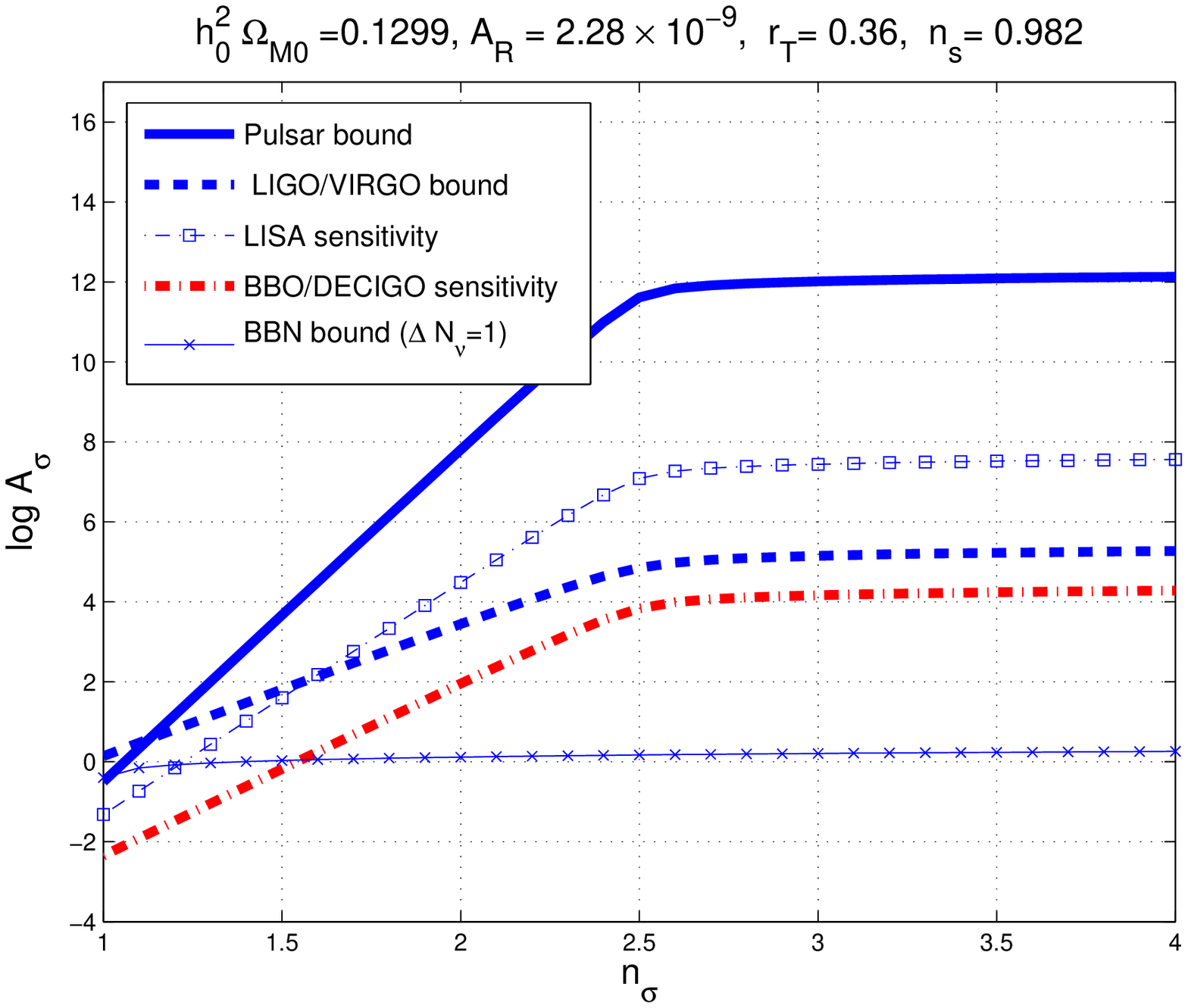}
\includegraphics[height=6cm]{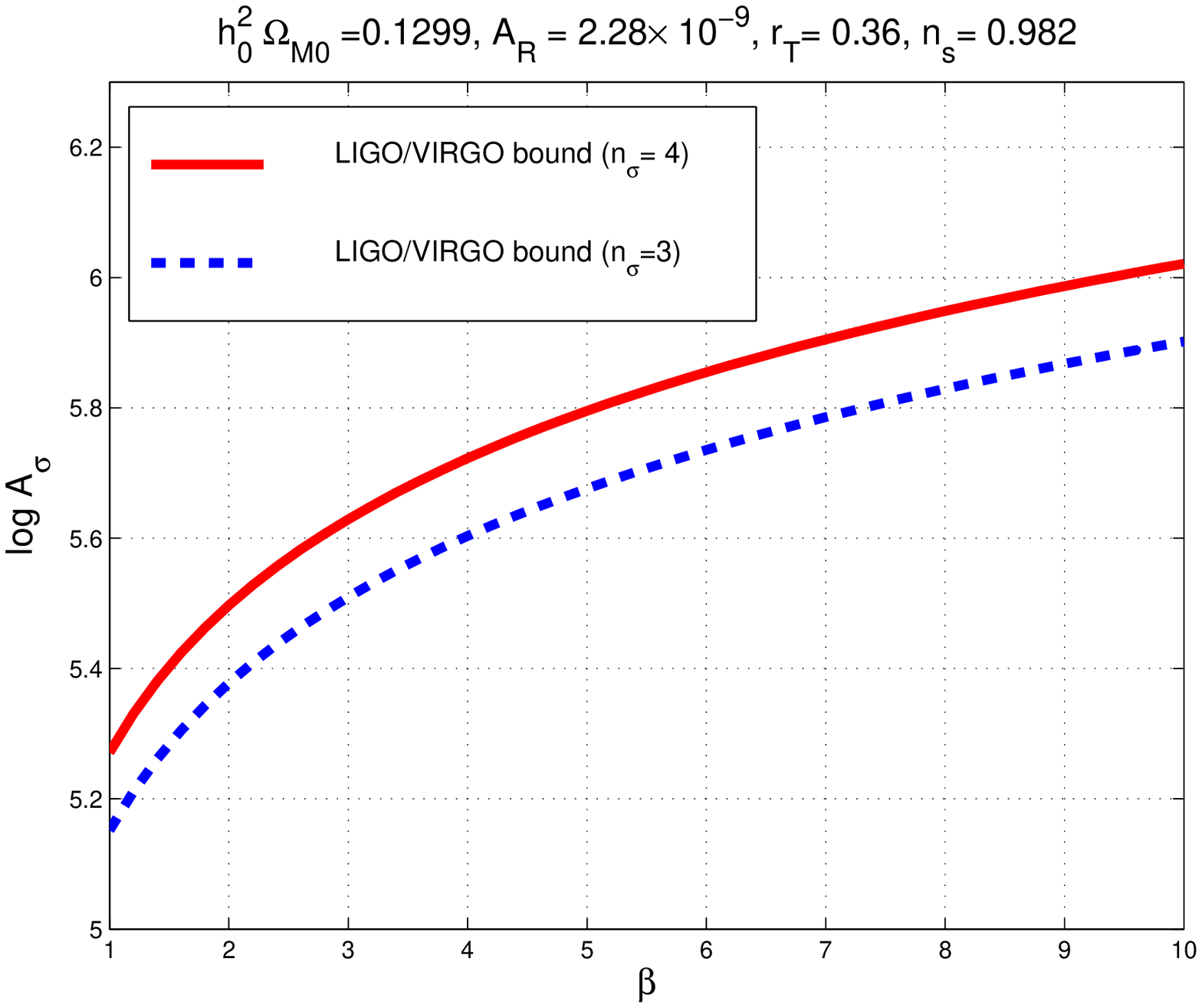}
\caption[a]{The pulsar and Ligo/Virgo bounds as well as 
the expected sensitivities of planned space-borne interferometers 
such as Lisa and Bbo/Decigo. As in Fig. \ref{figure2} $A_{\sigma}$ is expressed in units 
of the Planck mass $\overline{M}_{\mathrm{P}}$.}
\label{figure3}      
\end{figure}
In Fig. \ref{figure3} the current Ligo/Virgo bounds  are contrasted 
with the expected sensitivities of some of the planned 
space-borne interferometers such as Lisa \cite{lisa} and the 
Bbo/Decigo \cite{BBODECIGO} projects.  In the Lisa case 
the plots have been obtained by assuming a nominal 
sensitivity of $10^{-11}$ (for the spectral energy density) and 
in a frequency range centered around the mHz, i.e. 
$\nu_{\mathrm{Lisa}} = 10^{-3}$ Hz. In the Bbo/Decigo case the sensitivity has been taken to be $10^{-15}$ and for a typical frequency range centered around the $\nu_{\mathrm{BBO}} = \mathrm{Hz}$.
Furthermore, concerning the results of Figs. \ref{figure2} and \ref{figure3} the following comments are in order:
\begin{itemize}
\item{} the pulsar bound, because of the smallness of its frequency 
range, is not 
directly relevant to constrain the secondary contribution: in Planck units  the 
result would be  $A_{\sigma} < 10^{12}$ for $n_{\sigma} > 2$ (see Fig. 
\ref{figure3}, plot at the left); 
\item{} the Ligo/Virgo bound is more relevant insofar as it implies $A_{\sigma} < 10^{5.6} $ in Planck units for $n_{\sigma} > 3$: still since 
the frequency is too small and the corresponding window too narrow the relevance of the obtained constraint
is not comparable with the nucleosynthesis bound (see Figs. \ref{figure2} and
\ref{figure3}, plots at the left in both figures); 
\item{} the Ligo/Virgo bound is moderately effective in constraining 
the maximal frequency of the secondary contribution for fixed 
values of $n_{\sigma}$: from Fig. \ref{figure3} (plot at the right)  
$\beta < {\mathcal O}(6)$ implies $A_{\sigma} < {\mathcal O}(10^{5})$ for the case of violet spectral indices;
\item{} the big-bang nucleosynthesis bound, as expected, is the 
the most effective in constraining both $A_{\sigma}$ as well as 
$\beta$ in the interesting ranges of spectral indices; in particular 
the upper limit on $A_{\sigma}$ is ${\mathcal O}(10^{0.3})$ for 
$n_{\sigma} > 2$ (see Fig. \ref{figure2}).
\end{itemize}
By looking at Fig. \ref{figure3} (plot at the left) it is possible 
to spot a corner of the parameter space where $A_{\sigma}$ is 
compatible with the nucleosynthesis bound and where, simultaneously,
the putative signal could be detected by the forthcoming space-borne 
interferometers. This corner of the parameter space is, roughly 
speaking, for $1< n_{\sigma} < 2$ and $10^{-2} <A_{\sigma}< 1$. 
It should be however borne in mind that the results illustrated 
in Fig. \ref{figure3} depend on a specific value of $r_{\mathrm{T}}$.
Figures \ref{figure2} and \ref{figure3} present a scanning for different 
values of $\beta>1$. In the complementary case (i.e. $\beta <1$) the maximal 
frequency $\nu_{\mathrm{max}}$ (see Eq. (\ref{spect2a})) gets smaller. In spite 
of the latter occurrence the results discussed so far are only slightly modified 
in the interval $10^{-5} < \beta < 10$. Indeed the big-bang 
nucleosynthesis bound is sensitive to all the frequencies of the spectrum 
for $\nu>\nu_{\mathrm{bbn}}$ but a change in $\beta$ only changes $\nu_{\mathrm{max}}$ which is 
the ultraviolet cut-off buts also the normalization frequency of the waterfall spectrum. As far as 
the Ligo/Virgo bound is concerned, its relevance becomes slightly more pronounced 
for $\beta< 1$.  For instance for $10^{-4} < \beta < 10^{-3}$ the Ligo/Virgo bound would demand
$A_{\sigma} < 10^{3}$ in Planck units and in the case $n_{\sigma}=4$; for the same spectral index, Fig. \ref{figure3} 
would imply $A_{\sigma} < 10^{6}$ always in Planck units. 

Before closing the present section few final remarks are in order. 
The second-order contribution of the metric fluctuations can also 
produce a secondary background of relic gravitons whose 
spectral slope and amplitudes are, however, fixed by 
the slope and amplitude of the curvature perturbations (i.e. $n_{\mathrm{s}}$ and ${\mathcal A}_{{\mathcal R}}$) as discussed in \cite{assa}. Since 
$n_{\mathrm{s}} <1$ the resulting secondary spectra will contribute 
to smaller frequencies and will represent a (subleading) correction 
to the quasi-flat plateau of the primary contribution to the spectral energy density whose features will be briefly touched upon in section \ref{sec5}. 
Therefore the second-order contribution of the metric fluctuations has been neglected both because it is  proportional to ${\mathcal A}_{{\mathcal R}}^2$ 
and because it modifies preferentially the small frequencies.
 
Equations (\ref{WT23}) and (\ref{WT24}) do not account for the suppression of the quasi-flat plateau, at intermediate frequencies, due to the neutrino free streaming \cite{nu1}. The evolution of the relativistic degrees of freedom 
as well as the late dominance of dark energy have been neglected 
\cite{zh1}. 
As it will be explicitly shown in the next section in the combined effect 
of neutrino free streaming, of the late dominance of dark energy and 
of the evolution of the number of relativistic species affects 
the amplitude of the intermediate frequency range by, roughly, one order 
of magnitude and this is the reason why these effects, even if conceptually 
important, are often ignored in explicit analyses (see, e.g. \cite{efs}), given our ignorance 
of the value of $r_{\mathrm{T}}$.

\renewcommand{\theequation}{4.\arabic{equation}}
\setcounter{equation}{0}
\section{High-frequency gravitons in the concordance model}
\label{sec4}
The primary and the secondary contributions to the 
spectral energy density of the relic gravitons will now be illustrated 
using the numerical techniques described in \cite{mgs} (see also \cite{mgd1}) and enforcing the constraints derived in section \ref{sec3}. 
To avoid confusions, when talking about the {\em concordance model}
we refer to the {\em minimal extension} of the $\Lambda$CDM paradigm 
which includes tensors. The latter extension is realized, as swiftly mentioned 
after Eq. (\ref{WT24a}), when the spectral index does not run and 
$r_{\mathrm{T}}=16 \epsilon$ fixes, at once, the amplitude and the 
spectral index of the tensor power spectrum.
\begin{figure}[!ht]
\centering
\includegraphics[height=6cm]{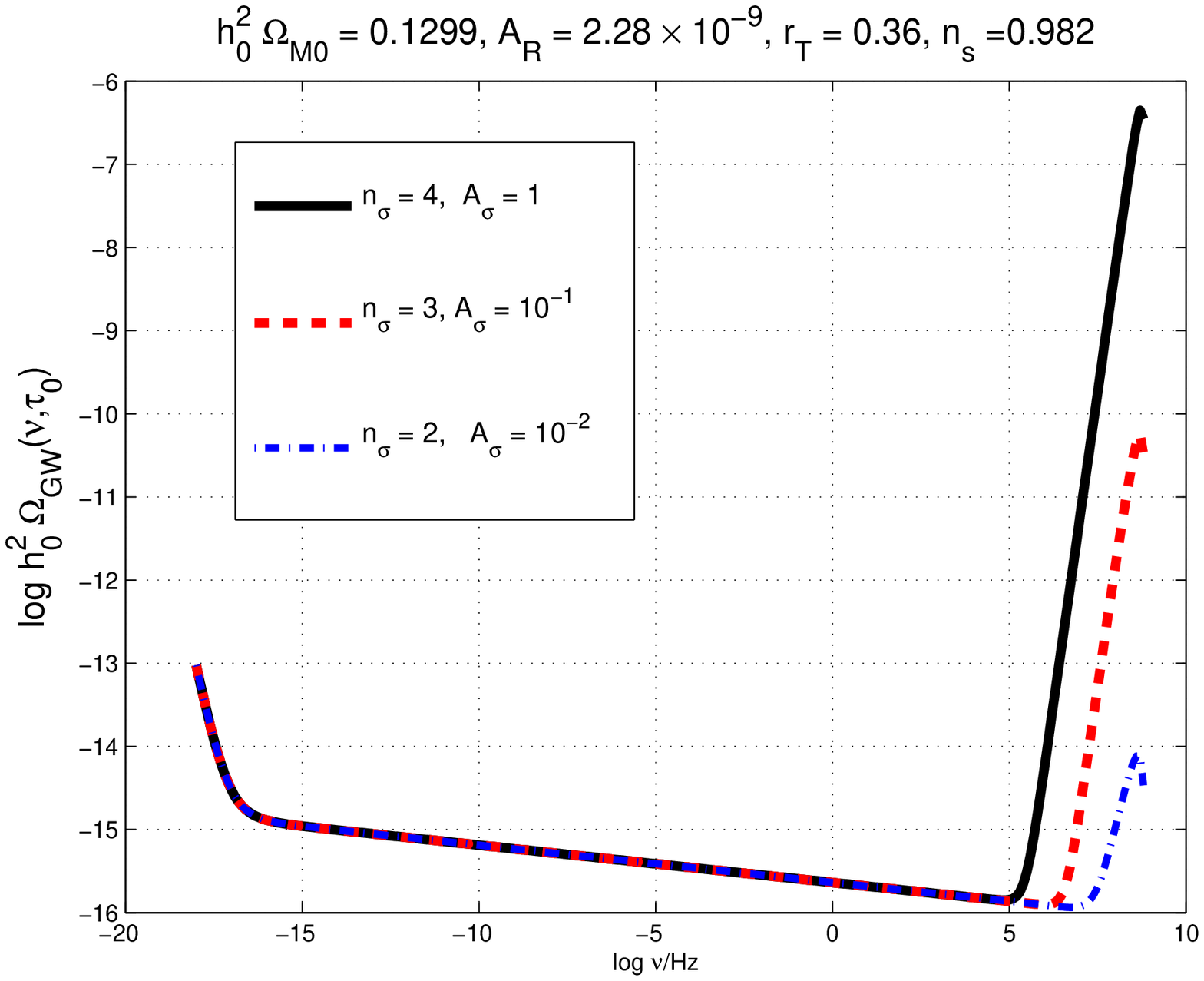}
\includegraphics[height=6cm]{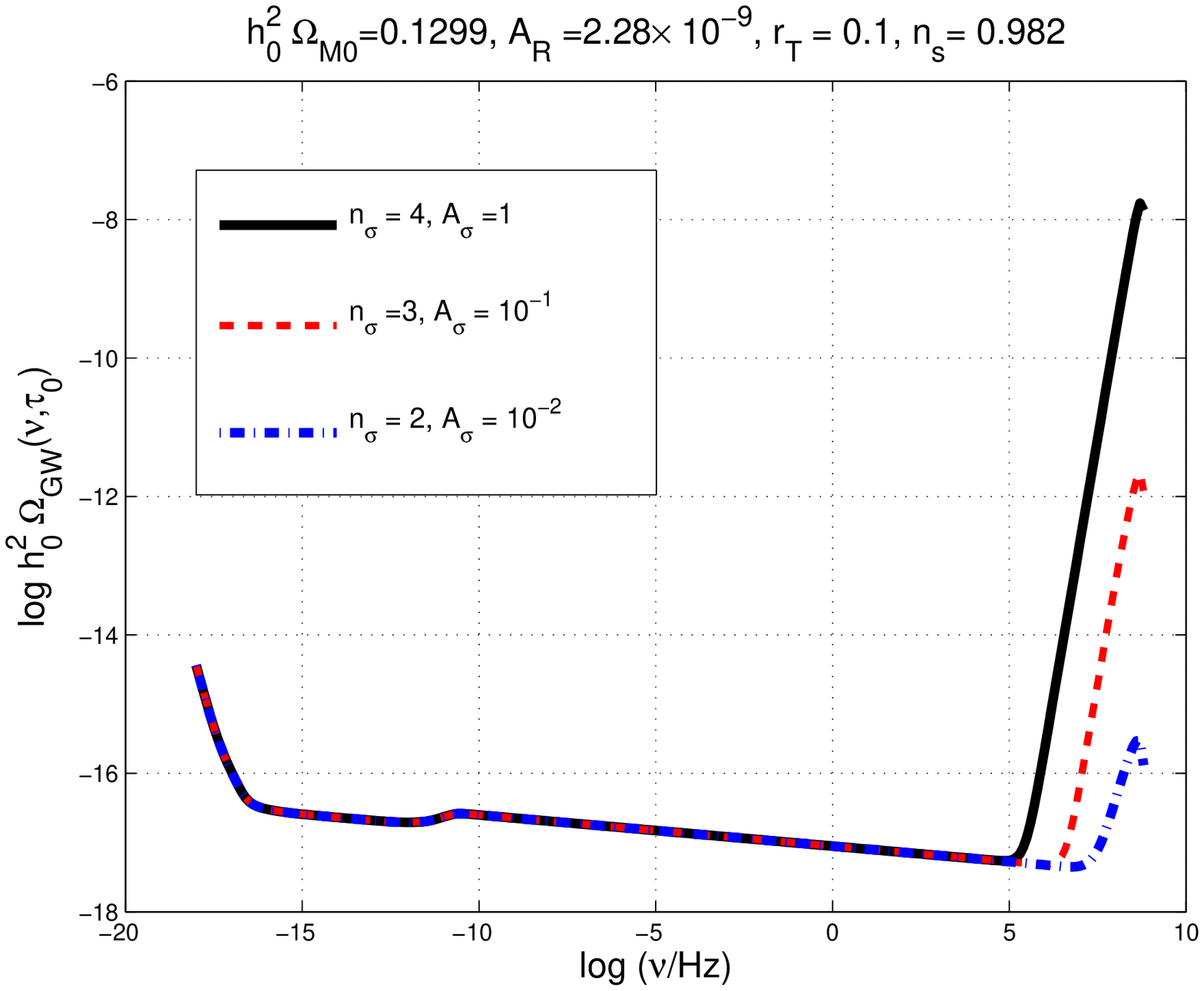}
\caption[a]{The primary and the secondary contributions to the 
spectral energy density are illustrated in the case of the parameters reported in Eq. (\ref{Par2}) (plot at the left) and in the case of the parameters 
of Eq. (\ref{Par2}) but taking into account also the neutrino free streaming and the late-time contribution of the dark energy.}
\label{figure4}      
\end{figure}
In Fig. \ref{figure4} the primary and secondary contributions are illustrated 
as they have been derived in the section \ref{sec3}. The maximal 
amplitude of $A_{\sigma}$ has been taken to be $1$ in Planck units. In the 
left plot of Fig. \ref{figure4} the effect of neutrino free streaming and 
the late-time dominance of the dark energy have been neglected. Conversely in the right plot both contributions have been included. 

The overall effect of the neutrino free streaming is to suppress the 
intermediate quasi-flat plateau for $\nu_{\mathrm{eq}} < \nu < \nu_{\mathrm{bbn}}$.  Indeed, assuming that the only collisionless 
species in the thermal history of the Universe are the neutrinos, the amount 
of suppression can be parametrized by the function \cite{nu1} (see also \cite{stein})
\begin{equation}
{\mathcal S}(R_{\nu}) = 1 -0.539 R_{\nu} + 0.134 R_{\nu}^2 
\label{ANIS3}
\end{equation}
where $R_{\nu}$ is the fraction of neutrinos in the radiation plasma, i.e. 
\begin{equation}
R_{\nu} = \frac{r}{r + 1}, \qquad r = 0.681 \biggl(\frac{N_{\nu}}{3}\biggr),\qquad R_{\gamma} + R_{\nu} = 1. 
\label{ANIS4}
\end{equation}
In the case $R_{\nu}=0$ (i.e. in the absence of collisionless patrticles) there is no suppression. If, on the contrary, 
$R_{\nu} \neq 0$ the suppression can even reach one order of magnitude. In the case $N_{\nu} = 3$, 
$R_{\nu} = 0.405$ and the suppression of the spectral energy density is proportional to ${\mathcal S}^2(0.405)= 0. 645$. 
The suppression due to neutrino free streaming will be effective for relatively 
small frequencies which are larger than $\nu_{\mathrm{eq}}$ and smaller than the  frequency corresponding to the Hubble radius at the time 
of big-bang nucleosynthesis, i.e. $\nu_{\mathrm{bbn}}$.

By carefully comparing the left and the right plots it is also possible to infer 
a global suppression which is actually due to the late-time dominance 
of dark energy. In the concordance model (i.e. the $\Lambda$CDM model)
the dark energy is parametrized in terms of a cosmological constant and, consequently, the redshift of $\Lambda$-dominance will be $1 + z_{\Lambda} \simeq (\Omega_{\Lambda}/\Omega_{\mathrm{M}0})^{1/3}$. The 
mode reentering the Hubble radius at $\tau_{\Lambda}$ will have 
a frequency 
\begin{equation}
 \nu_{\Lambda} = 2.629 \times 10^{-19}  \biggl(\frac{h_{0}}{0.735}\biggr) \biggl(\frac{\Omega_{\mathrm{M}0}}{0.243}\biggr)^{1/3} \biggl(\frac{\Omega_{\Lambda}}{0.757}\biggr)^{1/3} \,\, \mathrm{Hz}.
\label{LAM}
\end{equation}
The adiabatic damping of the mode function across $\tau_{\Lambda}$ reduces the amplitude of the spectral energy density by a factor 
$(\Omega_{\mathrm{M}0}/\Omega_{\Lambda})^2$. For the typical choice of parameters of Eq. (\ref{Par2}) the overall suppression of the spectral energy density is ${\mathcal O}(0.103)$. There is, in principle, a further modification 
of the spectral slope occurring between $0.3$ and $0.2$ aHz ($1$aHz = $10^{-18}$ Hz) which has been taken into account but which is 
quantitatively immaterial. 

Finally the overall suppression of the curves of Fig. \ref{figure4} receives also a (subleading)  contribution from the evolution of the number of (effective) relativistic degrees of freedom of the plasma entering 
the spectral energy density both as $g_{\rho}$ (i.e. the spin degrees of freedom pertaining to the total energy density) and as $g_{s}$ (i.e. 
the spin degrees of freedom pertaining to the total entropy density). 
In principle if a given mode $k$ reenters the Hubble radius at a temperature $T_{k}$ the spectral energy density 
of the relic gravitons is (kinematically) suppressed by a factor which can be written as (see, for instance, \cite{zh1})
 $(g_{\rho}(T_{k})/g_{\rho0})(g_{\mathrm{s}}(T_{k})/g_{\mathrm{s}0})^{-4/3}$.
At the present time  $g_{\rho0}= 3.36$ and $g_{\mathrm{s}0}= 3.90$.  The maximal expected suppression will be of the order of $0.38$ in the standard model where $g_{\rho} = g_{s} = 106.75$ for temperatures larger than the top quark mass. In popular supersymmetric extensions of the standard model $g_{\rho}$ and $g_{s}$  can be as high as, approximately, $230$. This will bring down the figure given above to $0.29$.

The spectra computed in the concordance model
the last remark is that the quasi-flat plateau has been computed 
long ago by different authors \cite{f1,f2,f3} and under different approximations. Thorough the years various completions of 
the standard paradigm have been proposed but none of them 
provides a strong gravitational wave background either at small or at 
large frequencies. An exception is represented by the presence 
of and early stiff phase \cite{mgb, ST1} which can arise also in braneworld 
scenarios \cite{ST2} as well as in other contexts \cite{ST3}. 
\begin{figure}[!ht]
\centering
\includegraphics[height=6cm]{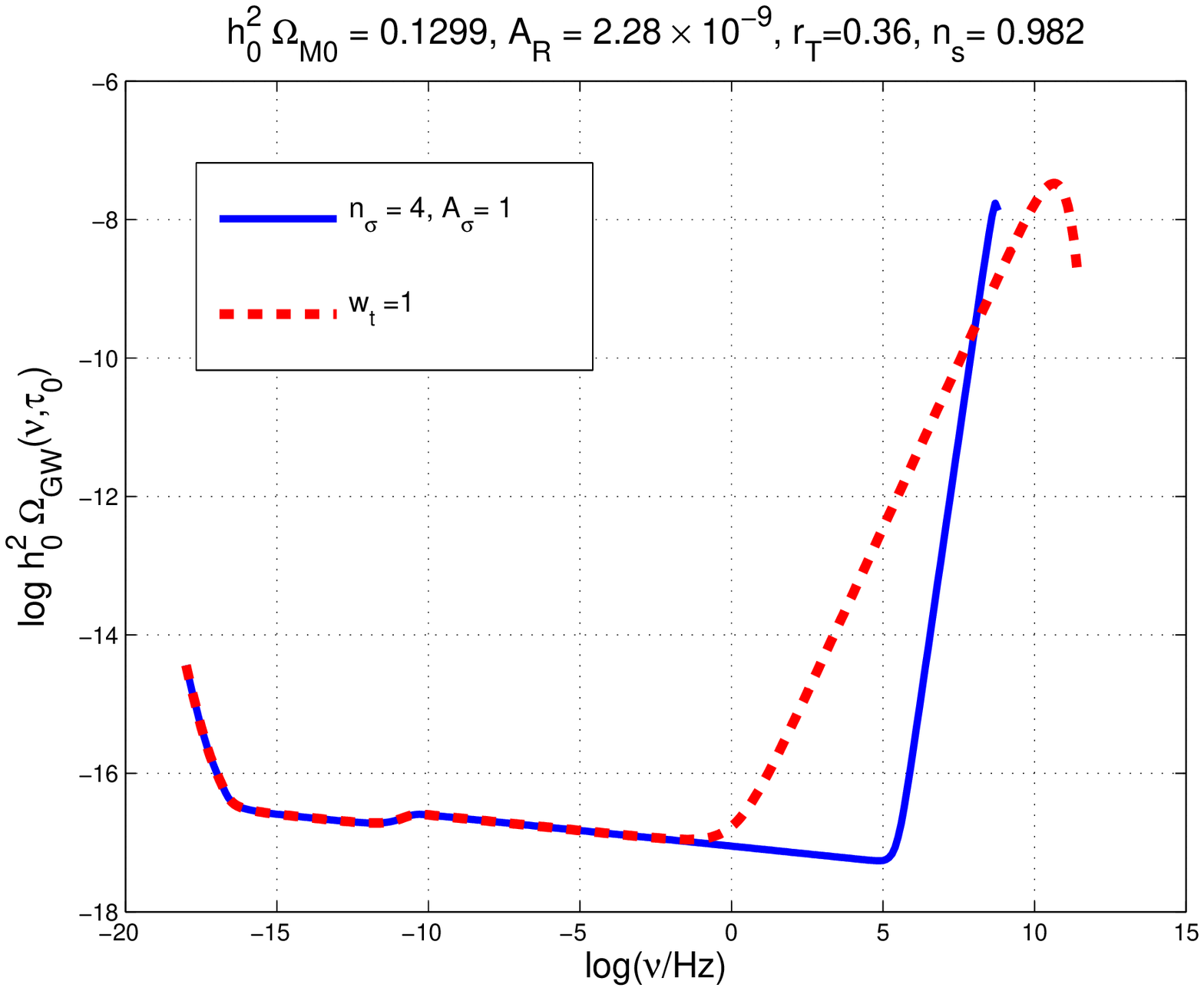}
\includegraphics[height=6cm]{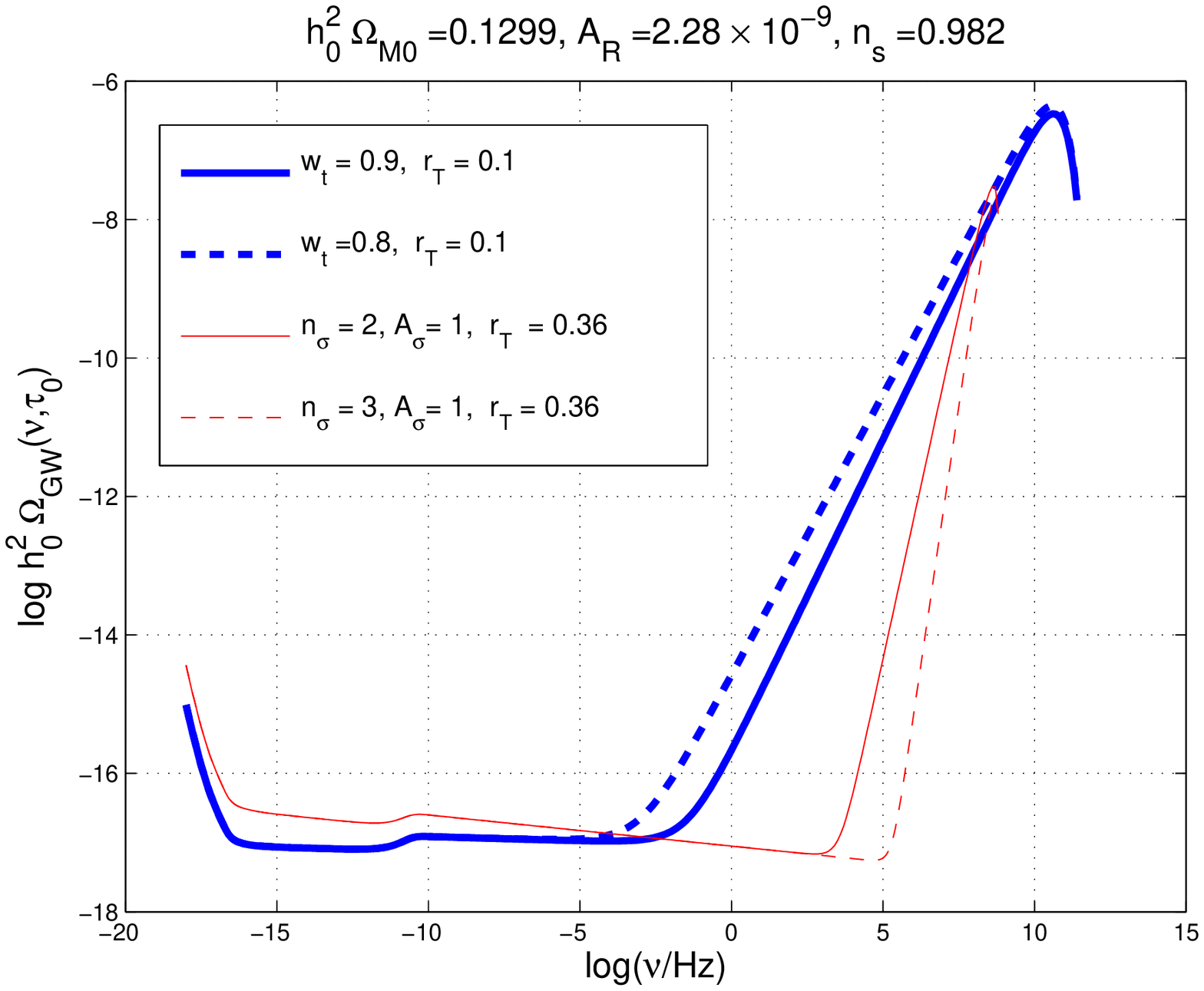}
\caption[a]{The secondary contribution to the spectral energy density of the relic gravitons is 
compared with the primary contribution computed in the T$\Lambda$CDM scenario, i.e. 
when the standard radiation-dominated phase is preceded by a stiff epoch characterized 
by a total speed of sound larger than the one 
of a radiation-dominated plasma. }
\label{figure5}      
\end{figure}
In Fig. \ref{figure5} the primary and secondary contributions 
to the relic graviton spectra in the concordance model is compared 
with the spectral energy density computed in the case of the 
what has been dubbed T$\Lambda$CDM model in \cite{mgd1}.
The T$\Lambda$CDM model (for tensor-$\Lambda$CDM) 
is an early time completion of the standard $\Lambda$CDM model 
where, after inflation, the dominance of radiation is 
delayed by a phase where the total speed of sound of the plasma 
is larger than the speed of sound of a radiation fluid. In terms 
of the barotropic index we shall have that, during the stiff phase, 
$w_{\mathrm{t}} > 1/3$. 
In Fig. \ref{figure5} (plot at the left) the case $w_{\mathrm{t}} =1$ is compared 
with the case typical of a waterfall field (i.e. $n_{\sigma} =4$) and for the same range of parameters of Eq. (\ref{Par2}). The case $w_{\mathrm{t}}=1$ 
corresponds to a total sound speed coinciding with the speed 
of light and the spectral energy density of relic gravitons induced in 
such a case have been discussed in various reprises \cite{mgb}.
For $ 1/3 < w_{\mathrm{t}} <1$ the situation is partially illustrated 
the right plot of Fig. \ref{figure5} where different values of $r_{\mathrm{T}}$
are also allowed. 

\renewcommand{\theequation}{5.\arabic{equation}}
\setcounter{equation}{0}
\section{Concluding remarks}
\label{sec5}
The secondary contribution to the spectral energy density of the relic gravitons induced by a waterfall-like field has been computed and 
constrained. The secondary contribution has been also combined with 
the standard signal arising in the case of the concordance $\Lambda$CDM paradigm and in its neighboring extensions such as the 
T$\Lambda$CDM model. 

The obtained results suggest the following series of considerations:
\begin{itemize}
\item{} for small frequencies (in the aHz range) the primary 
contribution of the concordance model remains unchanged by 
the secondary contribution;
\item{} for large frequencies, in the GHz range, the secondary contribution 
can be even $7$ or $8$ orders of magnitude larger than the 
nominal amplitude obtained, in the concordance model, 
when the tensor spectral index is not allowed to run and the 
WMAP 7yr constraints on $r_{\mathrm{T}}$ are implemented;
\item{} for intermediate frequencies, compatible with the 
observational window of wide-band interferometers the 
potential signal dies off faster than in the case of 
the T$\Lambda$CDM paradigm where a broad high-frequency 
spike arises because, right after inflation, the speed of sound  
is stiffer than the speed of sound of the radiation plasma;
\item{} the explicit constraints on the amplitude and on the spectral 
index of the waterfall field indicate that the nucleosynthesis 
constraint is more stringent than the direct Ligo/Virgo bounds as well as more stringent than all the other bounds usually applicable 
to relic graviton backgrounds;
\item{} the secondary contribution is still ineffective 
for frequencies coinciding either with the (putative) Lisa and Bbo/Decigo sensitivities, however, if the spectral index is sufficiently small (i.e. 
the high-frequency hump sufficiently broad) also the secondary contribution can be relevant also at intermediate frequencies by so overlapping 
with some of the predictions of the T$\Lambda$CDM model; this tuning 
is however not theoretically justified at present.
\end{itemize}
 The present results 
can be improved in a number of ways which have been partially 
mentioned in the bulk of the paper. In general terms 
steep spectra of waterfall-like fields might turn out to be 
a potential candidate for enhancing the spectral energy density 
of the relic gravitons at high frequencies.  Along a more conservative 
perspective the obtained results will contribute to a better quantitative 
understanding of the parameters characterizing waterfall-like fields 
in conventional and unconventional hybrid models. It is 
interesting to remark that if the concordance model is complemented 
by  a large tensor contribution at high-frequency low-frequency
determinations of $r_{\mathrm{T}}$ and high-frequency limits 
on the spectral energy density might not independent. In the 
future these kinds of models must be either ruled out (or ruled in) 
by combining low-frequency and high-frequency determinations 
of the spectral energy density.

\end{document}